\documentstyle[aps,epsf,axodraw,preprint,tighten]{revtex}

\begin{document}

\newcommand{\dfrac}[2]{\frac{\displaystyle #1}{\displaystyle #2}}
\draft
\preprint{VPI--IPPAP--00--05}

\title{Constraints on Gauged $B-3L_\tau$ and Related Theories}
\author{
Lay~Nam~Chang${}^{(1)}$\thanks{electronic address: laynam@vt.edu},
Oleg~Lebedev${}^{(1)}$\thanks{electronic address: olebedev@vt.edu},
Will~Loinaz${}^{(1,2)}$\thanks{electronic address: loinaz@alumni.princeton.edu}, and
Tatsu~Takeuchi${}^{(1)}$\thanks{electronic address: takeuchi@vt.edu}
}
\address{${}^{(1)}$Institute for Particle Physics and Astrophysics,
Physics Department, Virginia Tech, Blacksburg, VA 24061\\
${}^{(2)}$Department of Physics, Amherst College, Amherst MA 01002
}

\date{November 28, 2000}
\maketitle

\begin{abstract}
We consider extensions of the Standard Model with an extra
$U(1)$ gauge boson which couples to
$B-(\alpha L_e + \beta L_\mu + \gamma L_\tau)$
with $\alpha + \beta + \gamma = 3$. We show that
the extra gauge boson necessarily mixes with the $Z$, leading
to potentially significant corrections to the $Zf\bar{f}$ vertex.
The constraints on the size of this correction imposed by the
$Z$--pole data from LEP and SLD are derived.
\end{abstract}


\narrowtext

\section{Introduction}

A persistent mystery in particle physics today is
how nature distinguishes among the three generations of quarks and leptons 
and provides them with the observed mass hierarchy and mixings.
A popular approach in constructing a model which can potentially explain
this flavor problem is to extend the gauge symmetry of the
Standard Model (SM) and permit the three generations to
transform differently under the new symmetry.   
This idea is implemented, for instance, 
in topcolor \cite{Hill:1991at} 
and topcolor assisted technicolor \cite{Hill:1995hp}
models in which the third generation
transforms differently from the first two.

In extending the SM gauge group and assigning charges to the matter fields, 
care is needed to ensure anomaly cancellation.
However, extra care is necessary to further ensure 
\textit{charge orthogonality} when the extended gauge group contains 
multiple abelian factor groups \cite{Loinaz:1999qh}.  
Without charge orthogonality the abelian charges 
will mix kinetically \cite{Holdom:1986ag} 
under renormalization group running and the 
charge assignments lose scale--invariant meaning, 
rendering the model ill--defined.

In this paper, we examine these issues in the context of a series of
models introduced in Ref.~\cite{Ma:1998nq}. 
In these models, the SM gauge group is extended by an abelian factor to
$SU(3)_C \times SU(2)_L \times U(1)_Y \times U(1)_X$,
where the extra $U(1)_X$ gauge boson is coupled to some linear combination
of baryon and lepton flavor numbers\footnote{%
A model in which only the baryon number is gauged was considered in 
Ref.~\cite{Carone:1995aa}.}
which is anomaly free, {\it e.g.}
\[
X = B-3L_\tau,\; B-3L_e, \;B-\frac{3}{2}(L_\mu+L_\tau),\; {\rm etc.}
\]
These models were motivated by the desire to explain the masses and
mixing in the neutrino sector.
Unfortunately, the particle content of these models does not satisfy the
charge orthogonality condition (COC).
Consequently, the $X$ and $Y$ charges mix under renormalization group 
running and the model remains incomplete in the absence of a scale at 
which the charges are defined.
While one can always assume that the charges are those at the scale
at which the $U(1)_X$ symmetry breaks, thereby locking in the charges,
at higher energy scales the two $U(1)$ gauge bosons will 
couple to some scale--dependent linear combination of 
$X$ and $Y$ charges.  To avoid this one must go to the physical basis,
as discussed in Ref.~\cite{Loinaz:1999qh}, in which the charges are 
orthogonal and scale invariant.  However, the new scale invariant charges
can no longer be associated with the SM weak hypercharge or any
particular lepton flavor, and in general they will not even be rational.
It is therefore necessary that the COC be imposed as an additional
constraint if the initial charge assignment is to make any sense.

In the following,
we show below that the COC cannot be satisfied in this class of
models for {\it any} linear combination of $B$ and $L_{e,\mu,\tau}$
which is anomaly--free without the addition of extra matter fields.
Even when the COC is satisfied there can still be significant 
mixing between the $Z$ and the $X$, and this can show up in
$Z$--pole observables by breaking lepton universality.
We use the LEP and SLD $Z$--pole data to place significant constraints 
on the size of this mixing.
We find that, in the absence of additional sources of mixing,
the mass of the $X$ gauge boson is generically
required to be at least a few hundred GeV.

\section{Anomaly Cancellation and Charge Orthogonality}

Consider a model with the gauge group
$SU(3)_C \times SU(2)_L \times U(1)_Y \times U(1)_X$ where
$U(1)_Y$ is the putative weak hypercharge and $U(1)_X$ is an 
additional abelian factor group.
We choose the charge assignments of the quarks and leptons to be
\widetext
\begin{eqnarray*}
\left( \begin{array}{c} u_i \\ d_i \end{array} \right)_L
& \sim & \left(3,2,\dfrac{1}{6};\dfrac{1}{3}\right),
\qquad \; \; \; \;
u_{iR}
\;\sim\; \left(3,1,\dfrac{2}{3};\dfrac{1}{3}\right),\qquad
\; \;
d_{iR}
\;\sim\; \left(3,1,-\dfrac{1}{3};\dfrac{1}{3}\right), \cr
\left( \begin{array}{c} \nu_e \\ e \end{array} \right)_L
& \sim & \left(1,2,-\dfrac{1}{2};-\alpha\right),\qquad
e_{R}
\;\sim\; \left(1,1,-1;-\alpha\right),\qquad
\nu_{eR}
\;\sim\; \left(1,1, 0;-\alpha\right),\phantom{\dfrac{1}{2}} \cr
\left( \begin{array}{c} \nu_\mu \\ \mu \end{array} \right)_L
& \sim & \left(1,2,-\dfrac{1}{2};-\beta\right),\qquad
\mu_{R}
\;\sim\; \left(1,1,-1;-\beta\right),\qquad
\nu_{\mu R}
\;\sim\; \left(1,1, 0;-\beta\right),\phantom{\dfrac{1}{2}} \cr
\left( \begin{array}{c} \nu_\tau \\ \tau \end{array} \right)_L
& \sim & \left(1,2,-\dfrac{1}{2};-\gamma\right),\qquad
\tau_{R}
\;\sim\; \left(1,1,-1;-\gamma\right),\qquad
\nu_{\tau R}
\;\sim\; \left(1,1, 0;-\gamma\right),\phantom{\dfrac{1}{2}}
\end{eqnarray*}
\narrowtext
\noindent
where $i=1,2,3$ is the generation index, and $\alpha$, $\beta$, $\gamma$
are left arbitrary for the moment.  In effect, the $U(1)_X$ gauge boson
is chosen to couple to
\[
X = B - (\alpha L_e + \beta L_\mu + \gamma L_\tau).
\]
The addition of the right--handed neutrinos is necessary to make $U(1)_X$
a vectorial symmetry.
Note that even though we list three right--handed neutrinos,
any one whose $U(1)_X$ charge is chosen to be zero will effectively
decouple completely from the theory.
The minimal scalar sector necessary to break the gauge symmetry
into the usual $SU(3)_C\times U(1)_{\rm em}$ consists of the
regular Higgs doublet
\[
\left( \begin{array}{c} \phi^+ \\ \phi^0 \end{array} \right)
\sim \left(1,2,\frac{1}{2};0\right)
\]
and a neutral singlet
\[
\chi^0 \sim\left(1,1,0,\delta\right),\qquad \delta\neq 0,
\]
to break $U(1)_X$.
It is easy to show that anomaly cancellation leads to the condition
\begin{equation}
\alpha + \beta + \gamma = 3.
\label{anomaly}
\end{equation}

In non--GUT models with multiple gauged $U(1)$'s 
an additional constraint is necessary to ensure
that these groups do not mix through radiative corrections. As has been
discussed in Ref.~\cite{Loinaz:1999qh}, 
this requirement (at one loop) amounts to
\[
Tr[Q_X Q_Y]=0 \; ,
\]
the charge orthogonality condition.
In the model under consideration, the COC leads to the condition
\[
\alpha + \beta + \gamma = -1.
\]
Obviously, this and the anomaly cancellation condition Eq.~(\ref{anomaly})
cannot be satisfied simultaneously.
Therefore, the COC cannot be imposed 
regardless of the choice of $\alpha$, $\beta$, and $\gamma$,
and the model in its present form is ill--defined.

One way to rectify this problem is to change the charge assignments
in the minimal scalar sector so that the kinetic mixing due to the scalars 
cancels that due to the fermions.   However, this cannot be done so
easily since the scalar charges are fixed by the requirement that they
lead to the correct pattern of symmetry breaking, and also allow for
the necessary Yukawa couplings to give masses to the fermions.
Instead one may introduce new matter fields.
The set of new matter fields that are necessary to impose the COC
is not unique.   For instance, the COC can be imposed upon the 
fermion sector by an introduction of a pair of fermions with charge 
assignments given by
\begin{eqnarray*}
N_L & \sim & (1,1,a,b), \cr
N_R & \sim & (1,1,a,b),
\end{eqnarray*}
with $ab=-4$.
If the scalar sector is extended so that neutrino mixing can be generated,
the COC must be satisfied there also.
Therefore, the complete phenomenology of the model cannot be worked out
until all the extra fields have been specified.
One can nevertheless place a constraint on these models,
under a minimal set of assumptions, as we will discuss next.

\section{Vertex Corrections}

Even with the COC initially imposed at high energies, 
the $X$ and $Y$ charges will mix once some of the particles
decouple from renormalization group running.
This mixing can be fairly large at the $Z$ mass scale since it
will be proportional to $\ln(\Lambda/m_Z)$, where $\Lambda$ is the scale
at which decoupling occurs.   However, $X$--$Y$ mixing will lead to
the violation of lepton universality on the $Z$--pole which is 
well constrained by LEP and SLD data.   
In the following, we make the simplifying
assumption that all the non--SM particles decouple at
$M_X \gg m_Z$, the scale at which $U(1)_X$ breaks.  
Below $M_X$, only the SM particles survive decoupling and contribute
to $Z$--$X$ mixing.  Comparison of the size of this mixing with the
data will enable us to constrain $M_X$.

Consider possible corrections to the $Zf\bar{f}$ vertex at the
$Z$--pole.  There are two ways in which the $X$ boson can correct the
vertex.  The first is by dressing the vertex as shown in Fig.~1(a).
This correction leads to a shift in the $Zf\bar{f}$ couplings
given by:
\[
\frac{\delta h_{f_L}}{h_{f_L}} =
\frac{\delta h_{f_R}}{h_{f_R}} =
\frac{\alpha_X}{6\pi}(X_f^2)
\left(\frac{m_Z^2}{M_X^2}\ln\frac{M_X^2}{m_Z^2}\right),
\]
where $X_f$ is the $X$--charge of fermion $f$, and $\alpha_X =
g_X^2/4\pi$.
The second is through $Z$--$X$ mixing as shown in Fig.~1b.
Since charge orthogonality is broken below $M_X$ this correction
does not vanish and is given by
\widetext
\begin{equation}
\delta h_{f_L} =
\delta h_{f_R} =
-\frac{\alpha_X}{6\pi}(8s^2X_f)
\frac{m_Z^2}{M_X^2}
\left[ \ln\frac{M_X^2}{m_Z^2}
     + \dfrac{ \left(\frac{1}{2} - \frac{4}{3}s^2\right) }
             { 8 s^2 }
       \ln\frac{m_t^2}{m_Z^2}
\right]
\approx
-\frac{\alpha_X}{6\pi}(8s^2X_f)
\left(\frac{m_Z^2}{M_X^2}\ln\frac{M_X^2}{m_Z^2}\right),
\label{MixingCorrection}
\end{equation}
\narrowtext
\noindent
where $s^2$ is shorthand for $\sin^2\theta_W$, and the tree level
$Zf\bar{f}$ couplings are normalized as
\[
h_{f_L} = I_{3f} - Q_f s^2,\qquad
h_{f_R} =        - Q_f s^2.
\]
The top mass dependent term in Eq.~(\ref{MixingCorrection})
is due to the decoupling of the top quark below $m_t^2$,
but we will neglect it since it is suppressed compared to the other term.
Note that this correction is proportional to $s^2$ since
it is only the $U(1)_Y$ part of the $Z$ which mixes with the $X$.

Let us define
\[
\xi \equiv
\frac{\alpha_X}{6\pi}
\left(\frac{m_Z^2}{M_X^2}\ln\frac{M_X^2}{m_Z^2}\right).
\]
Then the sum of the dressing and mixing corrections can be written as
\begin{eqnarray*}
\delta h_{f_L} & = & \left( h_{f_L}X_f^2 - 8s^2 X_f \right) \xi, \cr
\delta h_{f_R} & = & \left( h_{f_R}X_f^2 - 8s^2 X_f \right) \xi.
\end{eqnarray*}
To place a constraint on the size of $\xi$ using
the data from precision electroweak measurements,
we follow the general procedure of Ref.~\cite{Takeuchi:1994zh}.
We assume that the only significant
non--SM vertex correction comes from $\xi$.
Since we will only use LEP and SLD observables which are
{\it ratios of coupling constants} in our analysis, oblique corrections
will only manifest themselves through a shift in the
the effective value of $\sin^2\theta_W$ \cite{Peskin:1990zt}.
We introduce the parameter $\delta s^2$ to account for this deviation:
\[
\sin^2\theta_W = [\sin^2\theta_W]_{\rm SM} + \delta s^2.
\]
We use $\delta s^2$ only as a fit parameter and extract no information
from it so that our results are independent of the Higgs mass.
The shifts in the left and right handed couplings are then
\begin{eqnarray*}
\delta h_{f_L} & = & -Q_f \delta s^2 + \left( h_{f_L}X_f^2 - 8s^2 X_f
\right) \xi, \cr
\delta h_{f_R} & = & -Q_f \delta s^2 + \left( h_{f_R}X_f^2 - 8s^2 X_f
\right) \xi.
\end{eqnarray*}

\section{Constraints from Precision Electroweak Measurements}

The shifts in the LEP/SLD observables due to the shifts in the coupling
constants are easily calculable.  For instance, the shift in the partial
decay width $Z\rightarrow f\bar{f}$ is given by
\begin{eqnarray*}
\frac{\delta\Gamma_f}{\Gamma_f}
& = & \frac{ 2 h_{f_L}\delta h_{f_L} + 2 h_{f_R}\delta h_{f_R} }
            { h_{f_L}^2 + h_{f_R}^2 } \cr
& = & -\frac{2( h_{f_L} + h_{f_R} )}{h_{f_L}^2 + h_{f_R}^2}\,Q_f\,\delta
s^2
\cr
&   & +\left[ 2X_f^2
             - 8s^2 X_f \frac{2( h_{f_L} + h_{f_R} )}{h_{f_L}^2 +
h_{f_R}^2}
       \right] \xi.
\end{eqnarray*}
Similary, the shift in the parity violating asymmetry
$A_f = (h_{f_L}^2-h_{f_R}^2)/(h_{f_L}^2+h_{f_R}^2)$ is given by
\begin{eqnarray*}
\frac{\delta A_f}{A_f}
& = & \frac{ 4 h_{f_L} h_{f_R} }{ (h_{f_L}^4 - h_{f_R}^4) }
      \left( h_{f_R} \delta h_{f_L} - h_{f_L} \delta h_{f_R} \right) \cr
& = & \frac{ 4 h_{f_L} h_{f_R} }
           { ( h_{f_L}^2 + h_{f_R}^2 )( h_{f_L} + h_{f_R} ) }
      \left[ Q_f\,\delta s^2 + 8s^2 X_f\,\xi \right]
\end{eqnarray*}
Note that the $X$ boson dressing correction which is proportional to
$X_f^2$ vanishes in $\delta A_f/A_f$.

The $\delta s^2$ and $\xi$ dependence of the observables we
use in our fit are as follows: 
\widetext
\begin{eqnarray}
\frac{\delta A_e}{A_e}
& = & -53.5\,\delta s^2 - 99.0\,\alpha\,\xi \cr
\frac{\delta A_\mu}{A_\mu}
& = & -53.5\,\delta s^2 - 99.0\,\beta\,\xi \cr
\frac{\delta A_\tau}{A_\tau}
& = & -53.5\,\delta s^2 - 99.0\,\gamma\,\xi \cr
\frac{\delta A_{\rm FB}(e)}{A_{\rm FB}(e)}
& = & 2\frac{\delta A_e}{A_e}
\;=\; -107\,\delta s^2 -198\,\alpha\,\xi \cr
\frac{\delta A_{\rm FB}(\mu)}{A_{\rm FB}(\mu)}
& = & \frac{\delta A_e}{A_e} + \frac{\delta A_\mu}{A_\mu}
\;=\; -107\,\delta s^2 -99.0\,(\alpha + \beta)\,\xi \cr
\frac{\delta A_{\rm FB}(\tau)}{A_{\rm FB}(\tau)}
& = & \frac{\delta A_e}{A_e} + \frac{\delta A_\tau}{A_\tau}
\;=\; -107\,\delta s^2 -99.0\,(\alpha + \gamma)\,\xi \cr
\frac{\delta R_e}{R_e}
& = & -0.840\,\delta s^2 +(1.18 + 1.09\,\alpha - 2\,\alpha^2)\,\xi
      +0.307\,\delta\alpha_s \cr
\frac{\delta R_\mu}{R_\mu}
& = & -0.840\,\delta s^2 +(1.18 + 1.09\,\beta - 2\,\beta^2)\,\xi
      +0.307\,\delta\alpha_s \cr
\frac{\delta R_\tau}{R_\tau}
& = & -0.840\,\delta s^2 +(1.18 + 1.09\,\gamma - 2\,\gamma^2)\,\xi
      +0.307\,\delta\alpha_s \cr
\frac{\delta \sigma^0_{\mathrm{had}}}{\sigma^0_{\mathrm{had}}}   
& = & 0.099\,\delta s^2 \cr
&&  + \left[ (1.599\,\alpha^2 - 2.006\,\alpha)
            -(0.401\,\beta^2  + 0.916\,\beta)
            -(0.401\,\gamma^2 + 0.916\,\gamma)
            -0.471
      \right] \xi \cr
&&  - 0.122\,\delta\alpha_s  \rule[-2mm]{0mm}{8mm}\cr
\frac{\delta R_b}{R_b} & = &  0.182\,\delta s^2 + 1.35\,\xi \cr
\frac{\delta R_c}{R_c} & = & -0.351\,\delta s^2 - 2.61\,\xi \cr
\frac{\delta A_{\rm FB}(b)}{A_{\rm FB}(b)}
& = & \frac{\delta A_b}{A_b} + \frac{\delta A_e}{A_e}
\;=\; -54.1\,\delta s^2 +(1.26 - 99.0\,\alpha)\,\xi \cr
\frac{\delta A_{\rm FB}(c)}{A_{\rm FB}(c)}
& = & \frac{\delta A_c}{A_c} + \frac{\delta A_e}{A_e}
\;=\; -58.7\,\delta s^2 -(4.80 + 99.0\,\alpha)\,\xi \cr
\frac{\delta A_b}{A_b} & = & -0.681\,\delta s^2 + 1.26\,\xi \cr
\frac{\delta A_c}{A_c} & = & -5.19\,\delta s^2 -4.80\,\xi
\label{FITCOEFS}
\end{eqnarray}
\narrowtext
We have assumed that the right--handed neutrinos are  heavy
and only the left--handed ones contribute to the invisible width of 
the $Z$.
The parameter $\delta\alpha_s$ gives the shift of the QCD coupling
constant $\alpha_s(m_Z)$ from its nominal value of 0.120:
\[  \alpha_s(m_Z) = 0.120 + \delta\alpha_s.  \]
Note that the correction from $Z$--$X$ mixing to the leptonic
asymmetry parameters $A_\ell$ ($\ell=e$, $\mu$, $\tau$) appears with
a large coefficient in Eq.~(\ref{FITCOEFS}).   This means that
lepton universality imposes a  strong constraint on $\xi$.

Here, we list the results of fitting the expressions in
Eq.~(\ref{FITCOEFS}) to the data listed in 
Table~\ref{LEP-SLD-DATA}\cite{LEP/SLD:2000} for two choices of $\alpha$, 
$\beta$, and $\gamma$ which were considered in Ref.~\cite{Ma:1998nq}.
In both cases $\alpha=0$, so we do not need to consider interference
between direct $Z$ and $X$ exchange.
The correlations among the data used are shown in Tables~\ref{LEPcorrelations}
and \ref{heavy-correlations}.

\begin{enumerate}

\item[(i)]
$\alpha=\beta=0$, $\gamma=3$ case:
\begin{eqnarray*}
\delta s^2     & = & -0.00067 \pm 0.00019 \cr
\xi            & = & \phantom{-}0.000015 \pm 0.000074 \cr
\delta\alpha_s & = & -0.0016 \pm 0.0032
\end{eqnarray*}
The correlation among the fit variables are shown in Table~\ref{corr003}
while the constraints from various observables in the $\delta s^2$--$\xi$
plane are shown in Fig.~\ref{fit003}.
The quality of the fit was $\chi^2 = 25.6/(19-3)$ with the largest
contributions coming from $A_{\rm FB}(b)$ (5.3) and
$\sigma^0_{\rm had}$ (3.6).
With such a large $\chi^2$, it is evident that including the
$X$--corrections do not improve the agreement between theory and
experiment.

To convert the limit on $\xi$ into a limit on $M_X$, we must assume a
value for $g_X$.   
For $g_X = g \approx 0.65$, the 
$1\sigma$ ($2\sigma$) upper bound on $\xi$ translates into:
\[
M_X \geq 860\;(580)\;{\rm GeV}.
\]
For $g_X = g' \approx 0.35$, the $1\sigma$ ($2\sigma$) bound is
\[
M_X \geq 370\;(220)\;{\rm GeV}.
\]
Interestingly enough, these bounds agree with that derived in 
Ref.~\cite{Ma:1998nq} which neglected both $Z$--$X$ mixing 
and oblique corrections.

\item[(ii)]
$\alpha=0$, $\beta=\gamma=1.5$ case:
\begin{eqnarray*}
\delta s^2     & = & -0.00063 \pm 0.00019 \cr
\xi            & = & -0.00008 \pm 0.00014 \cr
\delta\alpha_s & = & -0.0007 \pm 0.0034
\end{eqnarray*}
The correlation among the fit variables are shown in Table~\ref{corr01515}
while the constraints from various observables in the $\delta s^2$--$\xi$
plane are shown in Fig.~\ref{fit01515}.
The quality of the fit was $\chi^2 = 25.3/(19-3)$ with the largest
contributions coming from $A_{\rm FB}(b)$ (4.7) and
$A_{\rm LR}$ (3.3).  
For $g_X = g \approx 0.65$, the 
$1\sigma$ ($2\sigma$) upper bound on $\xi$ translates into:
\[
M_X \geq 1100\;(500)\;{\rm GeV}.
\]
For $g_X = g' \approx 0.35$, the $1\sigma$ bound is
\[
M_X \geq 490\;{\rm GeV}.
\]
The $2\sigma$ limit on $\xi$ does not lead to a 
constraint on $M_X$ in this case since
\[
\frac{m_Z^2}{M_X^2}\ln\frac{M_X^2}{m_Z^2} \le \frac{1}{e},
\]
with the maximum at
\[
\frac{m_Z^2}{M_X^2} = \frac{1}{e}.
\]
Since our analysis assumes $m_Z^2 \ll M_X^2$, the
approximation breaks down in the environs of this scale anyway,
invalidating any limits we may obtain.

\end{enumerate}

The limits for other choices of $\alpha$, $\beta$, and $\gamma$
are similar.  In Figs.~\ref{xi-limits} and \ref{mx-limits}, we plot the 
bounds on $\xi$ and $M_X$ for the $\alpha=0$ models as functions of 
$\beta = 3-\gamma$. 
As we can see, $M_X$ is generally required to be of the order of
a few hundred GeV.

\section{Discussion and Conclusions}

In this paper we have considered a class of models with an
abelian factor group of type $B-(\alpha L_e + \beta L_\mu + \gamma L_\tau)$ 
and have explored some of their phenomenological consequences.
In all cases considered, the quality of the fit to $Z$--pole 
electroweak observables is not improved over that of the Standard Model.
The new physics parameter $\xi$ violates lepton universality
and is strongly constrained by the leptonic observables.  The 
introduction of this parameter into the fit does not reconcile
the experimental values of $A_{\rm LR}$
and $A_{\rm FB}(b)$.  Thus the model does not provide a 
solution to the $A_b$ anomaly.  
We find that, under the assumption that 
the $X$ boson is heavier than the $Z$,
the $Z$--pole observables require that the mass of the extra gauge boson be
of order a few hundred GeV.

In this analysis we have considered only kinetic mixing 
of the $X$ and $Z$ bosons due to SM particles between the scales
$M_X$ and $m_Z$, but more complicated versions of this model are possible.  
For variants in which non--SM particles charged under the $U(1)$'s decouple 
above $M_X$ and cause the COC to be violated, the mixing will occur over a 
larger momentum range.
If the scalar sector of the model includes fields which transform
non--trivially under both $U(1)$'s, then their acquiring VEV's
can lead to mass mixing between the $X$ and the $Z$ \cite{Ma:1998nq}.
These additional sources of mixing may either dilute
or sharpen the constraints obtained here, but 
must be considered on a model by model basis since
they depend critically on the specifics of each model.

\section*{Acknowledgements}

This research is supported in part by the U.S.
Department of Energy, grant DE--FG05--92ER40709, Task A.

\newpage

\narrowtext

\begin{table}[h]
\begin{center}
\begin{tabular}{|c|c|c|}
Observable & Measured Value & ZFITTER Prediction \\
\hline\hline
\multicolumn{2}{|l|}{\underline{$Z$ lineshape variables}} & \\
$m_Z$                & $91.1872 \pm 0.0021$ GeV & input       \\
$\Gamma_Z$           & $2.4944 \pm 0.0024$ GeV  & unused      \\
$\sigma_{\rm had}^0$ & $41.544 \pm 0.037$ nb    & $41.474$ nb \\
$R_e$                & $20.803 \pm 0.049$       & $20.739$ \\
$R_\mu$              & $20.786 \pm 0.033$       & $20.739$ \\
$R_\tau$             & $20.764 \pm 0.045$       & $20.786$ \\
$A_{\rm FB}(e   )$   & $0.0145 \pm 0.0024$      & $0.0152$ \\
$A_{\rm FB}(\mu )$   & $0.0167 \pm 0.0013$      & $0.0152$ \\
$A_{\rm FB}(\tau)$   & $0.0188 \pm 0.0017$      & $0.0152$ \\
\hline
\multicolumn{2}{|l|}{\underline{$\tau$ polarization at LEP}} & \\
$A_e$        & $0.1483 \pm 0.0051$      & $0.1423$ \\ 
$A_\tau$     & $0.1425 \pm 0.0044$      & $0.1424$ \\
\hline
\multicolumn{2}{|l|}{\underline{SLD left--right asymmetries}} & \\
$A_{LR}$     & $0.15108 \pm 0.00218$    & $0.1423$ \\
$A_e$        & $0.1558 \pm 0.0064$    & $0.1423$ \\
$A_{\mu}$    & $0.137 \pm 0.016$    & $0.1423$ \\
$A_{\tau}$   & $0.142 \pm 0.016$    & $0.1424$ \\
\hline
\multicolumn{2}{|l|}{\underline{heavy quark flavor}}  & \\
$R_b$           & $0.21642 \pm 0.00073$ & $0.21583$ \\
$R_c$           & $0.1674  \pm 0.0038$  & $0.1722$ \\
$A_{\rm FB}(b)$ & $0.0988  \pm 0.0020$  & $0.0997$ \\
$A_{\rm FB}(c)$ & $0.0692  \pm 0.0037$  & $0.0711$ \\
$A_b$           & $0.911   \pm 0.025$   & $0.934$ \\
$A_c$           & $0.630   \pm 0.026$   & $0.666$ \\
\end{tabular}
\caption{LEP/SLD observables 
and their Standard Model predictions.
All data is from Ref.~\protect\cite{LEP/SLD:2000}.
The Standard Model predictions were calculated using ZFITTER v.6.21 
\protect\cite{ZFITTER:99} with default flag settings and
$m_t = 174.3$~GeV \protect\cite{CDF/D0:2000},
$m_H = 300$~GeV, and $\alpha_s(m_Z) = 0.120$ as input.}
\label{LEP-SLD-DATA}
\end{center}
\end{table}

\medskip

\widetext

\begin{table}[h]
\begin{center}
\begin{tabular}{|c|ccccccccc|}
& $m_Z$     & $\Gamma_Z$     & $\sigma_{\rm had}^0$
& $R_e$     & $R_\mu$     & $R_\tau$ 
& $A_{\rm FB}(e)$ & $A_{\rm FB}(\mu)$ & $A_{\rm FB}(\tau)$ \\
\hline
$m_Z$ 
& $1.000$            & $-0.008$           & $-0.050$ 
& $\phantom{-}0.073$ & $\phantom{-}0.001$ & $\phantom{-}0.002$ 
& $-0.015$           & $\phantom{-}0.046$ & $\phantom{-}0.034$ \\
$\Gamma_Z$
&                    & $\phantom{-}1.000$ & $-0.284$ 
& $-0.006$           & $\phantom{-}0.008$ & $\phantom{-}0.000$ 
& $-0.002$           & $\phantom{-}0.002$ & $-0.003$           \\
$\sigma_{\rm had}^0$
&                    &                    & $\phantom{-}1.000$ 
& $\phantom{-}0.109$ & $\phantom{-}0.137$ & $\phantom{-}0.100$ 
& $\phantom{-}0.008$ & $\phantom{-}0.001$ & $\phantom{-}0.007$ \\
$R_e$
&                    &                    & 
& $\phantom{-}1.000$ & $\phantom{-}0.070$ & $\phantom{-}0.044$ 
& $-0.356$           & $\phantom{-}0.023$ & $\phantom{-}0.016$ \\
$R_\mu$
&                    &                    & 
&                    & $\phantom{-}1.000$ & $\phantom{-}0.072$ 
& $\phantom{-}0.005$ & $\phantom{-}0.006$ & $\phantom{-}0.004$ \\
$R_\tau$ 
&                    &                    &
&                    &                    & $\phantom{-}1.000$ 
& $\phantom{-}0.003$ & $-0.003$           & $\phantom{-}0.010$ \\
$A_{\rm FB}(e)$ 
&                    &                    &
&                    &                    & 
& $\phantom{-}1.000$ & $-0.026$           & $-0.020$ \\
$A_{\rm FB}(\mu)$ 
&                    &                    & 
&                    &                    & 
&                    & $\phantom{-}1.000$ & $\phantom{-}0.045$ \\
$A_{\rm FB}(\tau)$
&                    &                    & 
&                    &                    & 
&                    &                    & $\phantom{-}1.000$ \\
\end{tabular}
\caption{The correlation of the $Z$ lineshape variables at LEP}
\label{LEPcorrelations}
\end{center}
\end{table}

\begin{table}[h]
\begin{center}
\begin{tabular}{|c|cccccc|}
& $R_b$              & $R_c$     
& $A_{\rm FB}(b)$    & $A_{\rm FB}(c)$     
& $A_b$              & $A_c$ \\
\hline
$R_b$ 
& $1.00$            & $-0.14$           & $-0.03$ 
& $\phantom{-}0.01$ & $-0.03$           & $\phantom{-}0.02$ \\
$R_c$
&                   & $\phantom{-}1.00$ & $\phantom{-}0.05$ 
& $-0.05$           & $\phantom{-}0.02$ & $-0.02$ \\
$A_{\rm FB}(b)$
&                   &                   & $\phantom{-}1.00$
& $\phantom{-}0.09$ & $\phantom{-}0.02$ & $\phantom{-}0.00$ \\
$A_{\rm FB}(c)$
&                   &                   &
& $\phantom{-}1.00$ & $-0.01$           & $\phantom{-}0.03$ \\
$A_b$
&                   &                   &
&                   & $\phantom{-}1.00$ & $\phantom{-}0.15$ \\ 
$A_c$
&                   &                   &
&                   &                   & $\phantom{-}1.00$ \\
\end{tabular}
\caption{The correlation of the heavy flavor variables from LEP/SLD.}
\label{heavy-correlations}
\end{center}
\end{table}

\narrowtext

\begin{table}[h]
\begin{center}
\begin{tabular}{|c|ccc|}
& $\delta s^2$  & $\xi$ & $\delta\alpha_s$ \\
\hline
$\delta s^2$ & $1.00$ & $-0.28$ & $\phantom{-}0.20$ \\ 
$\xi$        &        & $\phantom{-}1.00$ & $-0.23$ \\
$\delta\alpha_s$ &    &                   & $\phantom{-}1.00$ \\
\end{tabular}
\caption{The correlations of the fit variables for the $\alpha=\beta=0$,
$\gamma=3$ case.}
\label{corr003}
\end{center}
\end{table}

\narrowtext

\begin{table}[h]
\begin{center}
\begin{tabular}{|c|ccc|}
& $\delta s^2$  & $\xi$ & $\delta\alpha_s$ \\
\hline
$\delta s^2$ & $1.00$ & $-0.33$ & $\phantom{-}0.26$ \\ 
$\xi$        &        & $\phantom{-}1.00$ & $-0.39$ \\
$\delta\alpha_s$ &    &                   & $\phantom{-}1.00$ \\
\end{tabular}
\caption{The correlations of the fit variables for the $\alpha=0$,
$\beta=\gamma=1.5$ case.}
\label{corr01515}
\end{center}
\end{table}

\narrowtext

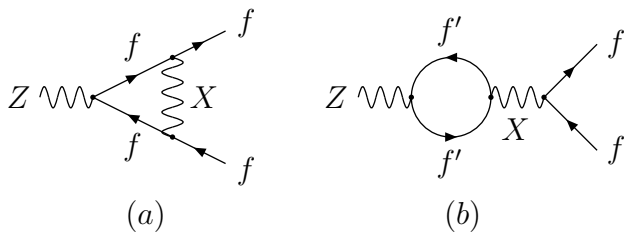
\begin{figure}[h]
\begin{center}
\begin{picture}(240,100)(0,-50)
\Vertex(30,0){1}
\Vertex(60,15){1}
\Vertex(60,-15){1}
\Photon(10,0)(30,0){4}{3}
\ArrowLine(30,0)(60,15)
\ArrowLine(60,15)(80,25)
\ArrowLine(80,-25)(60,-15)
\ArrowLine(60,-15)(30,0)
\Photon(60,-15)(60,15){4}{4}
\Text(2,0)[]{$Z$}
\Text(72,0)[]{$X$}
\Text(88,28)[]{$f$}
\Text(88,-28)[]{$f$}
\Text(45,19)[]{$f$}
\Text(45,-19)[]{$f$}
\Text(50,-45)[]{$(a)$}
\Vertex(150,0){1}
\Vertex(180,0){1}
\Vertex(200,0){1}
\Photon(130,0)(150,0){4}{3}
\Photon(180,0)(200,0){4}{3}
\ArrowArc(165,0)(15,0,180)
\ArrowArc(165,0)(15,180,0)
\ArrowLine(200,0)(220,20)
\ArrowLine(220,-20)(200,0)
\Text(122,0)[]{$Z$}
\Text(190,-12)[]{$X$}
\Text(228,20)[]{$f$}
\Text(228,-20)[]{$f$}
\Text(165,25)[]{$f'$}
\Text(165,-25)[]{$f'$}
\Text(170,-45)[]{$(b)$}
\end{picture}
\end{center}
\caption{One--loop vertex corrections to $Z \rightarrow f\bar{f}$.
Wavefunction renormalization corrections are not shown.} 
\label{FIG1}
\end{figure}

\newpage
\narrowtext

\begin{center}
\begin{figure}[h]
\begin{picture}(240,210)(0,0)
\unitlength=1mm
\put(25,30){$\sigma^0_{\rm had}$}
\put(24,50){$A_\tau({\rm LEP})$}
\put(66,46){$R_\tau$}
\put(26,19){$A_{\rm LR}$}
\put(56,19){$A_{\rm FB}(b)$}
\epsfbox[0 0 240 210]{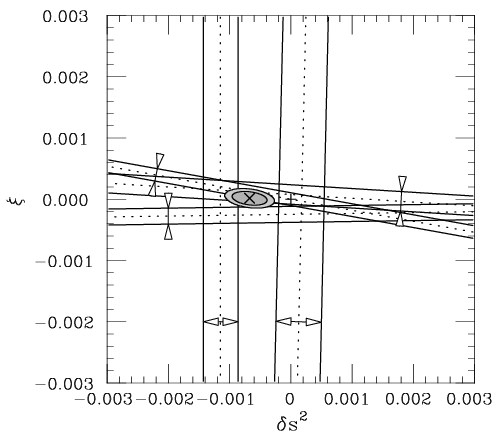}
\end{picture}
\caption{The 68\% and 90\% confidence contours in the
$\delta s^2$--$\xi$ plane for the $\alpha=\beta=0$,
$\gamma=3$ case.
The $1\sigma$ bounds from the observables leading to the
strongest constraints are also shown.
}
\label{fit003}
\end{figure}
\end{center}

\narrowtext

\begin{center}
\begin{figure}[h]
\begin{picture}(240,210)(0,0)
\unitlength=1mm
\put(26,28){$\sigma^0_{\rm had}$}
\put(24,55){$A_\tau({\rm LEP})$}
\put(65,17){$A_{\rm FB}(\mu)$}
\put(26,19){$A_{\rm LR}$}
\put(57,60){$A_{\rm FB}(b)$}
\epsfbox[0 0 240 210]{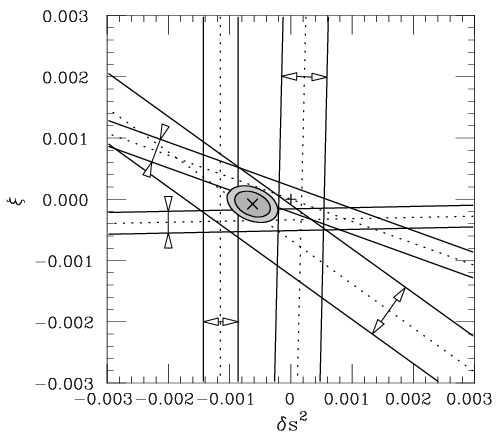}
\end{picture}
\caption{The 68\% and 90\% confidence contours in the
$\delta s^2$--$\xi$ plane for the $\alpha=0$,
$\beta=\gamma=1.5$ case.
The $1\sigma$ bounds from the observables leading to the
strongest constraints are also shown.
}
\label{fit01515}
\end{figure}
\end{center}

\narrowtext

\begin{center}
\begin{figure}[h]
\begin{picture}(240,170)(0,0)
\epsfbox[0 0 240 175]{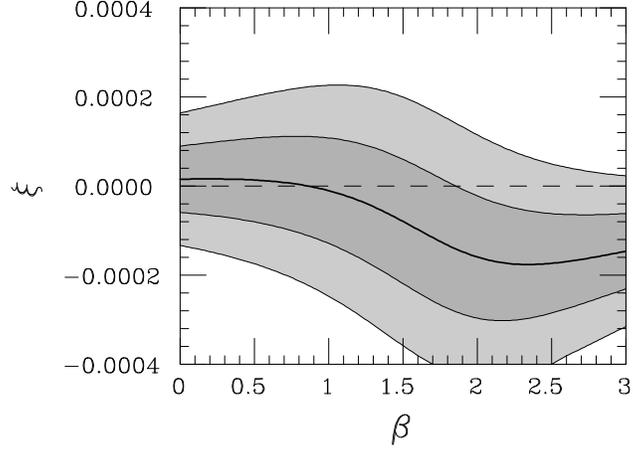}
\end{picture}
\caption{The $1\sigma$ (dark gray) and $2\sigma$ (light gray) limits on
$\xi$ for $\alpha=0$, $\beta+\gamma=3$ models.
}
\label{xi-limits}
\end{figure}
\end{center}

\narrowtext

\begin{center}
\begin{figure}[h]
\begin{picture}(240,170)(0,0)
\epsfbox[0 0 240 175]{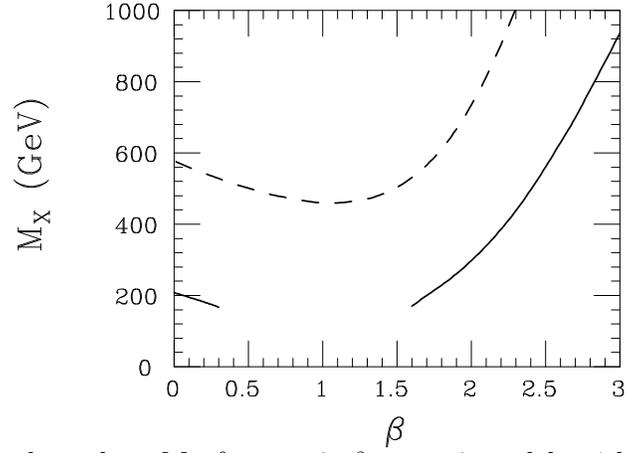}
\end{picture}
\caption{The $2\sigma$ lower bound on $M_X$ for 
$\alpha=0$, $\beta+\gamma=3$ models with
$g_X=0.65$ (dashed line) and $g_X=0.35$ (solid line).
No bound exists for the $g_X=0.35$ case between
$\beta\approx 0.4$ and $\beta\approx 1.5$.
}
\label{mx-limits}
\end{figure}
\end{center}

\newpage
\narrowtext

%
\end{document}